\documentclass{moriond}

\def\be{\begin{equation}}
\def\ee{\end{equation}}
\def\bea{\begin{eqnarray}}
\def\eea{\end{eqnarray}}

\usepackage{amssymb}
\usepackage{xspace}
\newcommand{\noun}[1]{\textsc{#1}}
\newcommand{\Pythia}{\noun{Pythia~8}\xspace}
\newcommand{\Matrix}{\textsc{Matrix}\xspace}
\newcommand{\minnlops}{\textsc{MiNNLO\textsubscript{PS}}\xspace}
\newcommand{\minloprime}{\noun{MiNLO$^\prime$}\xspace}
\newcommand{\geneva}{\noun{Geneva}\xspace}
\newcommand{\unnlops}{\noun{Unnlops}\xspace}
\newcommand{\powheg}{{\ttfamily POWHEG}\xspace}
\newcommand{\nnnllp}{N$^3$LL$^\prime$\xspace}
\newcommand{\pt}{p_{t}}
\newcommand{\kt}{k_{t}}
\newcommand{\ptj}{p_{t,j}}

\newcommand{\ptH}{p_{\sss\rm T,H}}
\newcommand{\ptZ}{p_{\sss\rm T,\ell\ell}}
\newcommand{\ptell}{p_{\sss\rm T,\ell^+}}
\newcommand{\mathd}{\mathrm{d}}
\def\sss{\scriptscriptstyle}
\newcommand\as{\alpha_{\sss\rm S}}
\newcommand{\abar}{\frac{\as}{2\pi}}
\newcommand{\abarmu}[1]{\frac{\as(#1)}{2\pi}}
\newcommand\PhiB{\Phi_{\scriptscriptstyle \rm F}}
\newcommand\PhiBJ{\Phi_{\scriptscriptstyle \rm FJ}}
\newcommand{\Fcorr}{F^{\sss\rm {corr}}}
\newcommand{\LambdaPWG}{\Lambda_{\rm pwg}}
\newcommand{\ptrad}{{p_{t,\rm rad}}}

\newcommand{\Photo}{}

\begin{document}
\vspace*{4cm}
\title{\minnlops: a new method to match NNLO QCD with parton showers}

\author{Emanuele RE}

\address{LAPTh, Universit\'e Grenoble Alpes, Universit\'e Savoie Mont Blanc, CNRS, F-74940 Annecy, France}

\maketitle\abstracts{
  I describe \minnlops, a novel method to match NNLO QCD
  computations with Parton Showers, and present a selection of results
  for color-singlet production at the LHC.
}

\section{Introduction}

The current experimental precision reached by the LHC experimental
collaborations (and even more the future prospects) requires
theoretical predictions whose formal accuracy go beyond the
computation of NLO QCD corrections and their matching to parton
showers (NLO+PS). Adding (N)NNLO QCD corrections, and NLO EW ones (or
combination thereof), is crucial, as often it is only through such
predictions that a comparison between data and theory is made possible
without being limited by large theoretical uncertainties, or by
missing shapes and normalization effects due to NNLO QCD and NLO EW
corrections. Nowadays the NNLO QCD corrections to essentially all the
$2\to 2$ processes~\footnote{As far as this talk is concerned, NNLO
  QCD corrections to off-shell diboson production with exact decays
  belong to this category} at the LHC are known, and NLO EW
corrections can be computed also for process with spectacularly high
multiplicity ($2\to 8$).

It is known that, in the corners of phase space where an hierarchy
between two or more scales develop, fixed-order QCD predictions fail,
due to the presence of large logarithms, that need to be resummed to
all orders. The recent progress in this field has been remarkable as
well, not only due to the extremely precise predictions obtained for
specific observables (in some cases, \nnnllp), but also because of the
development of parton-shower algorithms whose logarithmic accuracy can
be formally established for a wide class of observables, and
systematically improved.

A fully-differential Monte Carlo event generator that incorporates,
consistently, several of the above developments does not exist
yet. Nevertheless, the issue of matching NNLO QCD corrections to PS
has been already addressed by different groups, and NNLO+PS results
have been obtained with four methods: ``reweighted
\minloprime''~\cite{Hamilton:2012rf,Hamilton:2013fea},
\geneva~\cite{Alioli:2012fc,Alioli:2013hqa},
\unnlops~\cite{Hoeche:2014aia},
\minnlops~\cite{Monni:2019whf,Monni:2020nks}.  All the processes with
2 massless colored legs at LO can be described with this accuracy, and
many results have been already
obtained~\cite{Hamilton:2013fea,Karlberg:2014qua,Astill:2016hpa,Astill:2018ivh,Re:2018vac,Bizon:2019tfo,Alioli:2015toa,Alioli:2019qzz,Alioli:2020fzf,Alioli:2020qrd,Alioli:2021qbf,Alioli:2021egp,Cridge:2021hfr,Hoeche:2014aia,Hoche:2014dla,Hoche:2018gti,Monni:2019whf,Monni:2020nks,Lombardi:2020wju,Lombardi:2021rvg,Hu:2021rkt}.

In this contribution, I present
\minnlops\cite{Monni:2019whf,Monni:2020nks}, a new method to match
NNLO QCD computations with parton showers. I focus on color-singlet
production processes, although the method I describe has been already
successfully extended to deal with a colored and massive final state,
and used to simulate top-pair production at NNLO+PS accuracy,
obtaining, for the first time, a NNLO+PS result for a process with
more than two colored legs at
LO~\cite{Mazzitelli:2020jio}~\footnote{Results for top-pair production
  at NNLO+PS accuracy have been presented in a talk at this
  conference as well~\cite{ttbar_moriond}.}.

Denoting with $F$ the color-singlet final state, and with $\pt$ its
transverse momentum, the requirements of NNLO+PS prediction are:
\begin{itemize}
\item NNLO accuracy for observables that are inclusive on radiation (\emph{e.g.} $d\sigma/d y_F$).
\item NLO(LO) accuracy for $F+1$$(2)$ jet observables in the hard region
  (\emph{e.g.} $d\sigma/d\ptj$), possibly with a scale choice that is
  appropriate for each kinematic regime.
\item Resummation of soft/collinear logarithms where relevant
  (\emph{e.g.} $\sigma(\ptj<p_{t,veto})$ or $d\sigma/d\pt$).
\end{itemize}
The formal logarithmic accuracy that can be obtained by NNLO+PS
methods for specific observables is currently an open question. As far
as this manuscript is concerned, I will just assume that a NNLO+PS
prediction must preserve the logarithmic accuracy of the PS, that I will
assume to be leading-logarithmic (LL).
Theoretical ideas that already go beyond the NNLO+PS requirements
outlined above exist as well~\cite{Frederix:2015fyz,Prestel:2021vww}.

\section{The \minnlops method}
\subsection{\minnlops in a nutshell}
From $\pt$-resummation techniques~\footnote{For the case at hand, we
  use the direct-space resummation formalism of
  Ref.~\cite{Monni:2016ktx,Bizon:2017rah} ({\tt Radish}).}
one can show that the differential cross-section for $F+X$
production can be written as
\begin{equation}
  \label{eq:master1}
     \frac{\mathd\sigma}{\mathd\pt \mathd\PhiB} = \frac{\mathd}{\mathd\pt} \Big{\{}  \mathcal{L}(\Phi_{\rm F},\pt) \exp[-\tilde{S}(\pt)] \Big{\}} +  R_{\rm finite}(\pt)\,,
\end{equation}
where the luminosity $\mathcal{L}(\PhiB,\pt)$ contains the hard
virtual corrections and the coefficient functions for $F$ production
(expanded in powers of $\as(\pt)$), $\exp[-\tilde{S}(\pt)]$ is the
Sudakov form factor for the (direct-space) resummation of logarithms $\log(Q/\pt)$,
and $R_{\rm finite}(\pt)$ is the finite part of the $F+1$ jet
differential cross section $\mathd\sigma_{\sss\rm
  FJ}/\mathd\PhiBJ$. If $\mathcal{L}$ contains the hard-virtual and
the collinear coefficient functions up to second order in $\as(\pt)$,
and $R_{\rm finite}(\pt)$ is NLO accurate in the $\pt$ spectrum, the
integral over $\pt$ of Eq.~\ref{eq:master1} yields NNLO accurate
results for $F$ production.  The \minnlops method relies on the
observation that, after neglecting terms that, upon integration in
$\pt$, give contributions beyond the required accuracy,
Eq.~\ref{eq:master1} can be recast to a form that matches the \powheg
$\bar{B}$ function for the process $F+1$ jet:
\begin{eqnarray}
  \frac{\mathd\bar{B}(\PhiBJ)}{\mathd\PhiBJ}  &=&
  \exp[-\tilde{S}(\pt)]\bigg\{
      {\abarmu{\pt}\left[\frac{\mathd\sigma_{\scriptscriptstyle\rm FJ}}{\mathd\PhiBJ}\right]^{(1)} \left(1+\abarmu{\pt} [\tilde{S}(\pt)]^{(1)}\right)} \nonumber\\
      &~& {+ \left(\abarmu{\pt}\right)^2\left[\frac{\mathd\sigma_{\scriptscriptstyle\rm FJ}}{\mathd\PhiBJ}\right]^{(2)} } + {\left(\abarmu{\pt}\right)^3 [D(\PhiB,\pt)]^{(3)} {\Fcorr(\PhiBJ)}} \bigg\}\,,
      \label{eq:master2}
\end{eqnarray}
where 
\begin{equation}
       D(\PhiB,\pt) = -\frac{\mathd \tilde{S}(\pt)}{\mathd \pt} {\cal L}(\PhiB,\pt)+\frac{\mathd {\cal L}(\PhiB,\pt)}{\mathd \pt}\,,
\end{equation}
and $[X]^{(k)}$ is defined by $X=\sum_k (\abar)^k [X]^{(k)}$.

In Eq.~\ref{eq:master2} one recognizes the \minloprime formula, and
the extra term $[D(\pt)]^{(3)}$ needed to get NNLO accuracy.  As $D$
is extracted from equations valid in the $\pt\to 0$ limit, it formally
depends on $(\PhiB,\pt)$: in order to obtain the final expression of
Eq.~\ref{eq:master2} for an improved $\bar{B}$ function that depends,
instead, on $\PhiBJ$, one needs to introduce a smooth mapping and a
projection $\Fcorr(\PhiBJ)$, so that, when integrating the $D F$ term
in Eq.~\ref{eq:master2} over $\PhiBJ$ at fixed $(\PhiB,\pt)$, one
recovers exactly $D(\PhiB,\pt)$.

Truncating $D$ at the third order is formally correct, and
it is the approach used in ref.~\cite{Monni:2019whf}, although by keeping its full expression, \emph{i.e.}
\begin{equation}
  \left(\abarmu{\pt}\right)^3 [D(\pt)]^{(3)}  \to
  D -\abarmu{\pt} [D(\pt)]^{(1)}- \left(\abarmu{\pt}\right)^2[D(\pt)]^{(2)}\,,
\end{equation}
the dependence of the final results from higher-order subleading terms
(whose size can be numerically non-negligible) is minimized. In fact, by
keeping all the terms in $D$, the initial total derivative present in
Eq.~\ref{eq:master1} is preserved. The latter choice was introduced
in~\cite{Monni:2020nks}, and it was shown to yield better agreement between
NNLO+PS and NNLO results, as well as, together with other choices
discussed in the same paper, a more reliable scale variation band.

As a final comment, we notice that, starting from a point in the
$\PhiBJ$ phase space, the second radiation is generated using the
usual \powheg mechanism:
\begin{equation}
  \mathd\sigma={\bar B}(\PhiBJ)\ \mathd \PhiBJ\
  \bigg\{\Delta_{\rm pwg} (\LambdaPWG) + \mathd \Phi_{\sss\rm rad}
  \Delta_{\rm pwg} (\ptrad)  \frac{R (\PhiBJ{}, \Phi_{\sss\rm rad})}{B
    (\PhiBJ{})}\bigg\}\,,
\end{equation}
where $\Delta_{\rm pwg}$ is the \powheg Sudakov form factor governing
the second emission probability. If the first and the second emission
are strongly ordered, the above equation reproduces the emission
pattern of the first two emissions of a $\kt$-ordered parton shower,
thereby proving that the \minnlops method preserves the LL accuracy of
a $\kt$-ordered shower.

\subsection{Results}
In this section we show results for Drell-Yan and Higgs
production via gluon fusion. The NNLO results have been obtained with
\Matrix~\cite{Grazzini:2017mhc}, whereas the \minnlops ones are
obtained using the optimized approach described in
Ref.~\cite{Monni:2020nks}, after applying the
\Pythia parton shower, but without including hadronization and MPI
effects.

\begin{figure}[!htb]
  \begin{minipage}{0.5\linewidth}
    \centerline{\includegraphics[scale=0.35]{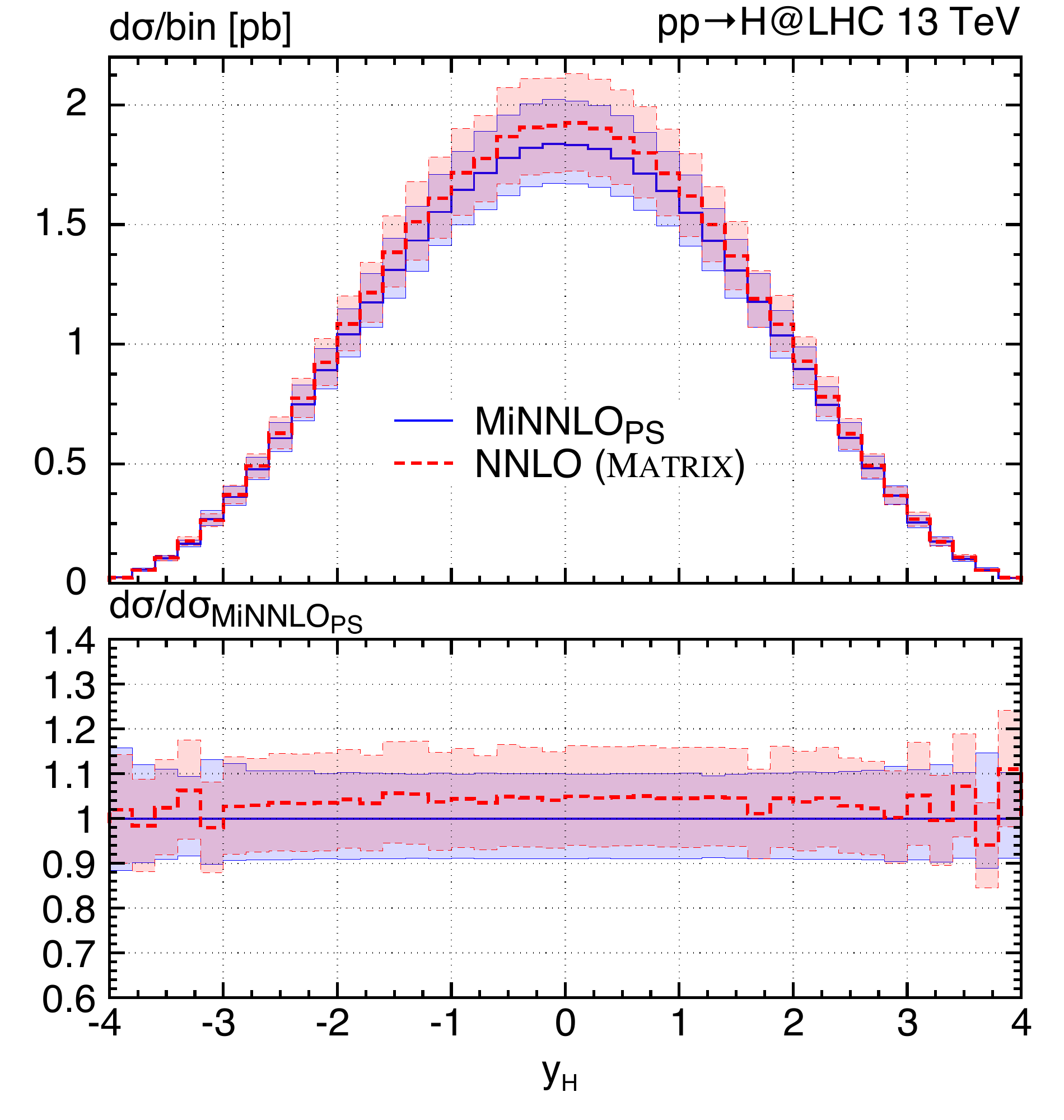}}
  \end{minipage}
  \begin{minipage}{0.5\linewidth}
    \centerline{\includegraphics[scale=0.35]{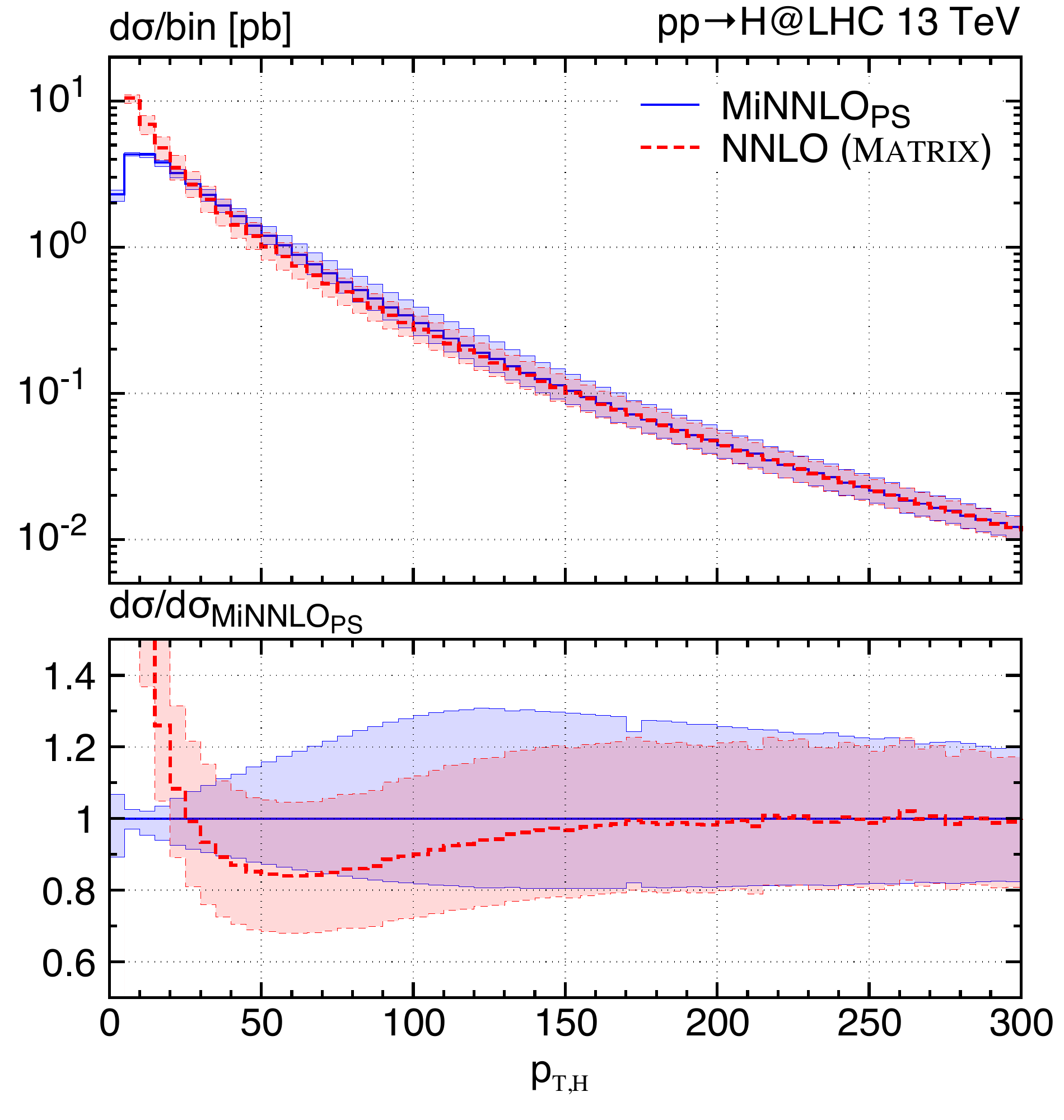}}
  \end{minipage}
  \caption{Comparison of NNLO (red) and \minnlops (blue) result for Higgs production via gluon fusion. The panel on the left (right) displays the Higgs boson rapidity (transverse momentum).}
  \label{fig:Higgs}
\end{figure}

In Fig.~\ref{fig:Higgs} we show predictions for the Higgs boson
rapidity (left) and its transverse momentum (right).
The agreement between the NNLO and the NNLO+PS prediction for the
Higgs rapidity is very good, differences are of the order of few
percent, the shape is predicted correctly and the size of the
theoretical uncertainty also agrees rather well between the two
predictions. For the Higgs transverse momentum, we observe good
agreement at large $\ptH$ values, as expected by the fact that the
\minnlops scale choice adopted in Ref.~\cite{Monni:2020nks} is such
that, at large $\ptH$, it approaches $M_H$, that is the scale used in
the NNLO result. At low values of $\ptH$, the typical Sudakov shape is
observed.

\begin{figure}[!htb]
  \begin{minipage}{0.5\linewidth}
    \centerline{\includegraphics[scale=0.35]{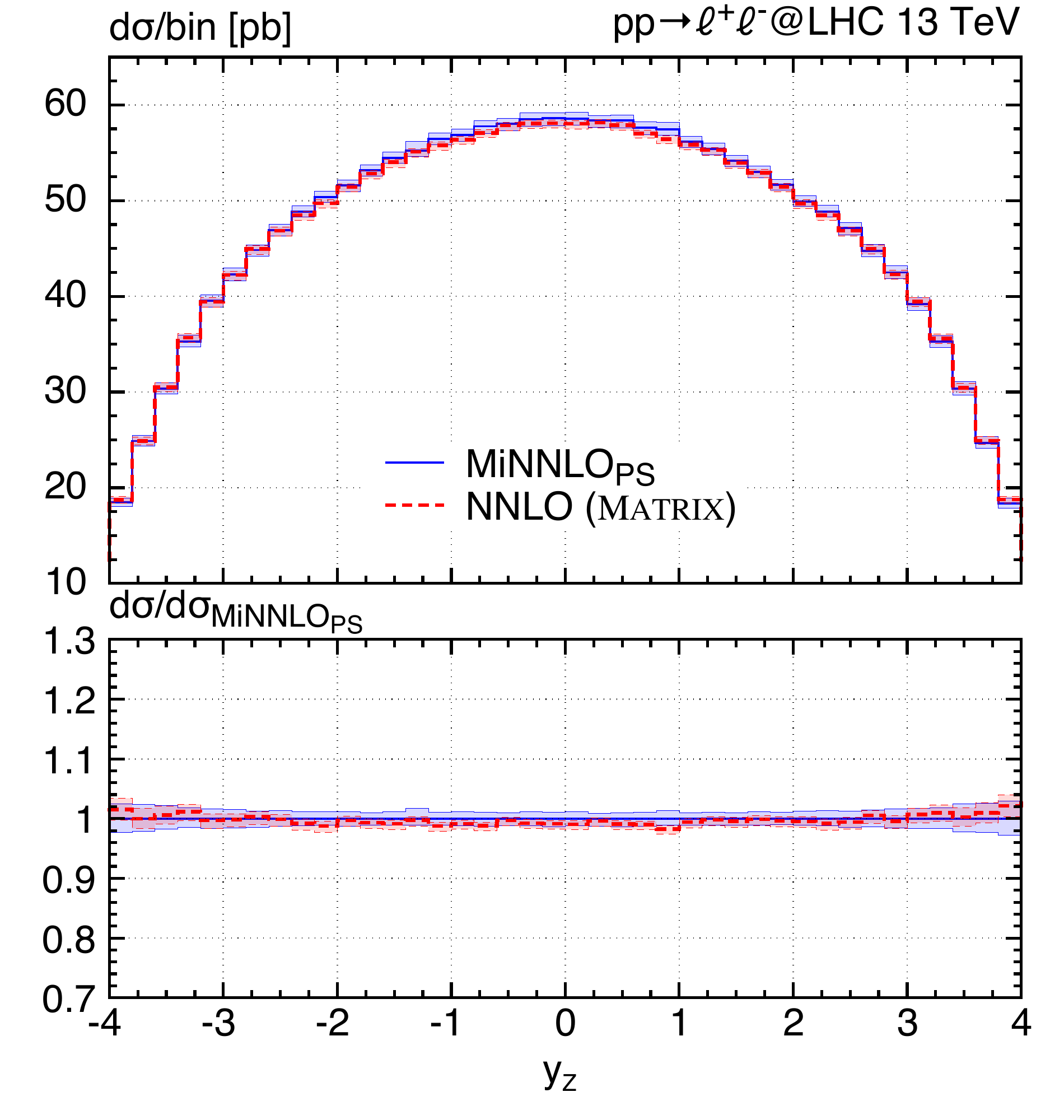}}
  \end{minipage}
  \begin{minipage}{0.5\linewidth}
    \centerline{\includegraphics[scale=0.35]{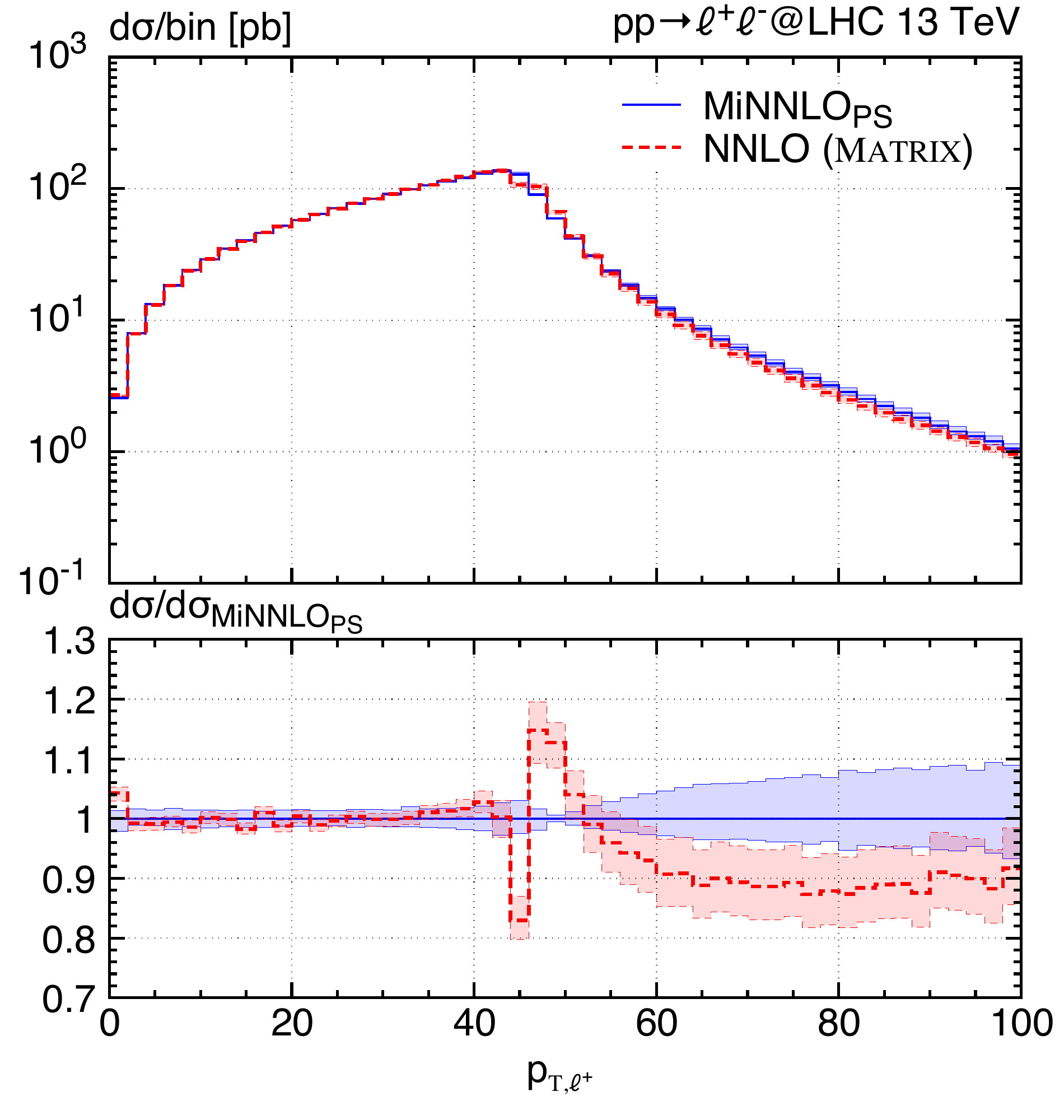}}
  \end{minipage}
  \caption{Comparison of NNLO (red) and \minnlops (blue) result for
    Drell-Yan production. The panel on the left (right) displays the
    dilepton rapidity (positron transverse momentum).}
  \label{fig:DY}
\end{figure}

In Fig.~\ref{fig:DY} we show instead the rapidity of the dilepton
system in Drell-Yan production (left) and the transverse momentum of
the positron (right).
The agreement of the results for the $Z$-boson rapidity is very
remarkable, and it also depends on the choice of the recoil scheme
adopted in \Pythia. The plot of the lepton transverse momentum shows
the expected features: an extremely precise prediction below the
Jacobian peak (with both results being NNLO accurate), a perturbative
instability of the NNLO result close to the Jacobian peak that gets
smooth by the PS, and a NLO-like uncertainty band of the two
results above the peak, with a different normalization due to the
different scale choices: for the NNLO result we have
$\mu=M_{\ell\ell}$, whereas in \minnlops one has $\mu\simeq \ptZ$, and
the region $\ptell\gtrsim M_Z/2$ is dominated by kinematic
configurations where the $Z$ boson has a small transverse momentum.

As mentioned in the introduction, results for
$Z\gamma$~\cite{Lombardi:2020wju} and $W^+W^-$~\cite{Lombardi:2021rvg}
production have been obtained recently with the \minnlops approach
too, as well as NNLO+PS predictions for $t\bar{t}$
production~\cite{Mazzitelli:2020jio}

\section*{Acknowledgments}
The results presented in this contribution have been obtained in
collaboration with P. F.~Monni, P.~Nason, M.~Wiesemann, and
G.~Zanderighi.

\end{document}